\documentclass[aps,prl,twocolumn,floatfix,superscriptaddress]{revtex4}

\usepackage{epsfig}
\usepackage{amsmath}

\begin{document}
\preprint{DUKE-TH-02-224-R}

\title{Light from Cascading Partons in Relativistic Heavy-Ion 
       Collisions}
\author{Steffen A.~Bass}
\affiliation{Department of Physics, Duke University, 
             Durham, NC 27708-0305}
\affiliation{RIKEN BNL Research Center, Brookhaven National Laboratory, 
	Upton, NY 11973, USA}
\author{Berndt M\"uller}
\affiliation{Department of Physics, Duke University, 
             Durham, NC 27708-0305} 
\author{Dinesh K.~Srivastava}
\affiliation{Physics Department, McGill University,
             3600 University Street, Montreal, H3A 2T8, Canada} 
\altaffiliation{on leave from: Variable Energy Cyclotron Centre, 
             1/AF Bidhan Nagar, Kolkata 700 064, India}            
\date{\today}
\begin{abstract}
We calculate the production of high energy photons from Compton
and annihilation processes as well as fragmentation off quarks in
the  Parton Cascade Model. The multiple scattering of partons is seen to 
lead to a substantial production of high energy photons, 
which rises further when parton multiplication due to 
final state radiation is included. The photon yield is found to be
directly proportional to the number of hard collisions and thus provides
valuable information on the pre-equilibrium reaction dynamics.
\end{abstract}
\maketitle
High-energy collisions of heavy nuclei are studied with the goal to 
explore the properties of matter at energy densities many times that
of normal nuclear matter. There are strong theoretical reasons to
believe that matter under these conditions, when thermally equilibrated,
forms a Quark-Gluon Plasma (QGP), a state of matter in which quarks
and gluons are not confined to hadrons. The discovery of this state
of matter and the study of its properties would constitute a major
advance in our understanding of quantum chromodynamics (QCD). 

 The parton 
cascade model (PCM) \cite{GM} was proposed to provide a detailed 
description of the temporal evolution of nuclear collisions at high 
energy, from the onset of hard interactions among the partons of the 
colliding nuclei up to the moment of hadronization. The PCM is based
on a relativistic Boltzmann equation for the time evolution of the
parton density in phase space due to perturbative QCD interactions
including scattering and radiation in the leading logarithmic 
approximation.

During the epoch where the partonic picture of sequential 
perturbative interactions among QCD quanta may be valid, important
many-body effects, such as parton shadowing, color screening, 
and Landau-Pomeranchuk-Migdal (LPM) suppression of gluon radiation,
complicate the single-particle dynamics of parton transport. Obviously,
the PCM involves the application of perturbative QCD to a hitherto
untested domain. 
 This poses the question - how  can one test the  correctness of this
approach?
This is 
clearly not possible by studying the hadronic final state, since
-- if the dense matter equilibrates --  
all information about its dynamics before equilibrium will be lost.
One may look for remnants of the pre-equilibrium dynamics in the
form of very energetic partons,
 which did not lose
a large fraction of their initial momentum due to interaction in the
dense medium and  produce jets.
 Even better suited to this purpose are weakly interacting 
probes of the pre-equilibrium dynamics, such as quanta that interact 
only electromagnetically: photons and leptons.

\begin{figure}[tb]
  \begin{center}  
  \epsfig{file=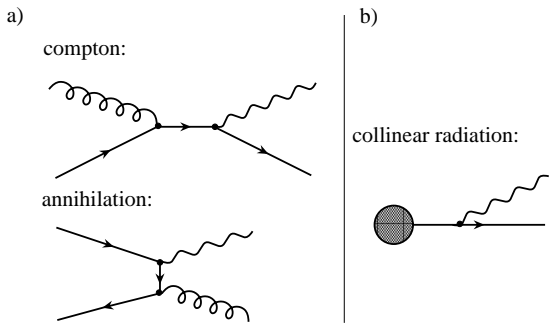,width=8.2cm}
  \caption{Photon production processes included in our calculation.} 
  \label{fig0}
  \end{center}
\end{figure}

Photons, which are produced from Compton ($q g \rightarrow q \gamma$), 
annihilation ($q \overline{q} \rightarrow g \gamma$), and 
bremsstrahlung ($q^\star \rightarrow q \gamma$) processes
 (see Fig.~\ref{fig0}), thus
assume considerable importance. These are analogous to the 
the processes governing the energy loss of energetic partons, 
where gluons are emitted instead of photons. Photons, however,  
once emitted, almost always escape the system, and their energy 
and momentum will remain unaffected~\cite{Fei:76} as they will only 
rarely interact with the surrounding medium. For this reason they 
are excellent and reliable probes of the evolution of the partonic 
cascade in nuclear collisions. In fact, one can argue that the
photons confirm the presence of parton cascading processes after
the initial primary parton-parton collisions.

In this letter we address several questions relating to the value
of photons as probes of the dense matter created in a nuclear 
collision: Are photons sensitive to the pre-equilibrium partonic
reaction dynamics, i.e. parton re-scattering? What effect do 
parton fragmentation processes have on photon production? Is there
a direct connection between the photon yield and the number of 
hard perturbative parton-parton collisions during the reaction?
In view of our specific focus, we do not consider the production 
of photons from a late-stage, thermal mixed phase or hadronic phase. 
These contributions are quite well understood \cite{joe}, and can
easily be added to our results.

The fundamental assumption underlying the PCM is that the state 
of the dense partonic system can be characterized by a set of 
one-body distribution functions $F_i(x^\mu,p^\alpha)$, where $i$
denotes the flavor index ($i = g,u,\bar{u},d,\bar{d},\ldots$)
and $x^\mu, p^\alpha$ are coordinates in the eight-dimensional
phase space. The partons are assumed to be on their mass shell,
except before the first scattering. In our numerical implementation, 
the GRV-HO parametrization \cite{grv} is used, and the parton 
distribution functions are sampled at an initialization scale 
$Q_0^2$ ($\approx (p_T^{\text{min}})^2 $; see later) to create a discrete 
set of particles. Partons generally propagate on-shell and along 
straight-line trajectories between interactions. Before their first 
collision, partons may have a space-like four-momentum, especially 
if they are assigned an ``intrinsic'' transverse momentum.  

The time-evolution of the parton distribution is governed by a 
relativistic Boltzmann equation:
\begin{equation}
p^\mu \frac{\partial}{\partial x^\mu} F_i(x,\vec p) = {\cal C}_i[F]
\label{eq03}
\end{equation}
where the collision term ${\cal C}_i$ is a nonlinear functional 
of the phase-space distribution function. The calculations discussed
below include all lowest-order QCD scattering processes between 
massless quarks and gluons \cite{Cutler.78}, as well as all 
($2 \to 2$) processes involving the emission of photons 
($qg \to q\gamma$, ${\bar q}g \to {\bar q}\gamma$, 
$q{\bar q} \to g\gamma$, $q{\bar q} \to \gamma\gamma$).
A low momentum-transfer cut-off $p_T^{\text{min}}$ is needed to 
regularize the infrared divergence of the perturbative parton-parton 
cross sections. A more detailed description of our implementation is 
in preparation \cite{bms_big1}.

We account for the final state radiation~\cite{Ben87} following a 
hard interaction in the parton shower approach. In the leading 
logarithmic approximation, this picture corresponds to a sequence 
of nearly collinear $1 \to 2$ branchings: $a \to bc$. Here $a$ is 
called the mother parton, and $b$ and $c$ are called daughters.
Each daughter can  branch again, generating a structure with a
tree-like topology. We include all such branchings allowed by
the strong and electromagnetic interactions.
The probability for a parton to branch is given in terms of the
variables $Q^2$ and $z$.  $Q^2$ is the momentum scale of the
branching and $z$ describes the distribution of the energy of the 
mother parton $a$ among the daughters $b$ and $c$, such that $b$ 
takes the fraction $z$ and $c$ the remaining fraction $1-z$. The 
differential probability to branch is:
\begin{equation}
{\cal P}_a=\sum_{b,c} \frac{\alpha_{abc}}{2\pi}P_{a \to bc}
           \frac{dQ^2}{Q^2} dz
\end{equation}
where the sum runs over all allowed branchings. The $\alpha_{abc}$ is 
$\alpha_{em}$ for branchings involving emission of a photon and 
$\alpha_s$ for the QCD branchings. The splitting kernels 
$P_{a \rightarrow bc}$ are given in \cite{Alt77}. The collinear 
singularities in the showers are regulated by terminating the 
branchings when the virtuality of the (time-like) partons drops 
to $\mu_0$, which we take as 1 GeV.  In principle, one could take 
a smaller value for the cut-off $\mu_0$ for a quark fragmenting 
into a photon~\cite{TS}, but we have not done so as 
we are only interested in high energy photons here. The soft-gluon 
interference is included as in \cite{Ben87}, namely by selecting 
the angular ordering of the emitted gluons.
An essential difference between emission of a photon and a parton
in these processes is that the parton encounters further interactions
and contributes to the build-up of the cascade, while the photon
almost always (in our approximation, always) leaves the system 
along with the information about the interaction.

Some of these aspects were discussed by authors of Ref.~\cite{SG}
and explored within framework of the PCM implementation VNI \cite{vni}. 
Unfortunately, several components of their numerical implementation were
incorrect. Our present work is based on a thoroughly revised and
extensively tested implementation of the parton cascade model by the 
present authors, called VNI/BMS \cite{bms_big1}.

\begin{figure}[tb]
  \begin{center}  
  \epsfig{file=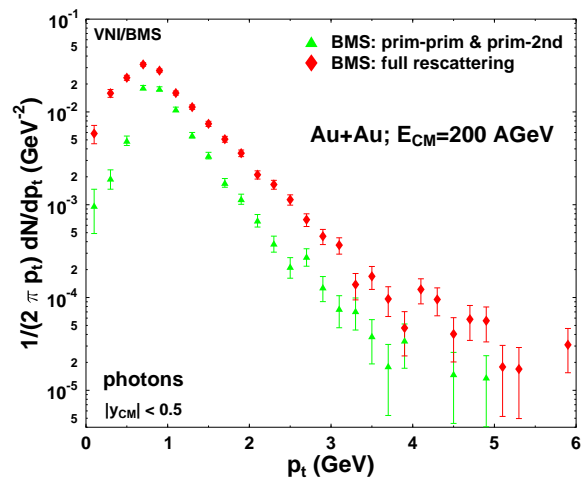,width=8.2cm}
  \caption{Transverse momentum distribution of photons from central
   collision of gold nuclei at RHIC (see text for an explanation of 
   the symbols).} 
  \label{fig1}
  \end{center}
\end{figure}

As a first step, we discuss the results obtained when the parton 
cascade is restricted to two-body scattering.  We are
considering central collisions ($b=0$) of Au nuclei at full RHIC 
energy ($\sqrt{s} = 200$ GeV per nucleon pair).  We have confirmed
the accuracy of our code by comparing the analytical pQCD prediction 
for prompt photon production, due to Compton and annihilation processes,
with our results, when we use an eikonal approximation in the cascade. 
Our results are shown in Fig.~\ref{fig1}, where we compare the 
contributions from interactions involving at least one primary parton 
(triangles) with the result obtained with full binary cascading (diamonds). 
Note that contribution from primary partons does not reflect the momentum 
broadening (Cronin effect) due to multiple soft scattering.  
The figure shows that most of the photons between 2 and 4 GeV 
have their origin in the multiple semi-hard scattering of partons.
This lends support to a recent prediction that there would be a 
substantial production of photons from the passage of high energy 
quarks through the QGP which may be produced in such secondary
collisions \cite{FMS.02}. 
We find the average squared transverse momentum of the
incoming rescattered partons at the photon vertex to be 
$\langle p_t^2 \rangle \approx 2.1$~(GeV/c)$^2$, confirming previous 
phenomenological analyses of nuclear broadening effects in prompt photon
production \cite{dumi01}.

\begin{figure}[tb]
  \begin{center}  
  \epsfig{file=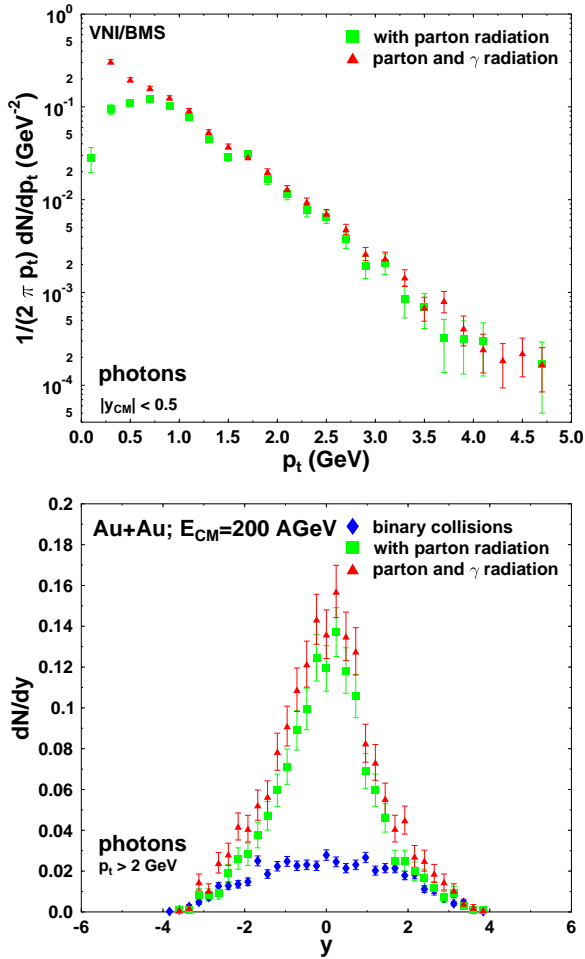,width=8.2cm}
  \caption{Upper frame: photon transverse momentum spectra for contributions
from collisions and fragmentations (triangles) and from collisions alone
(squares), when parton fragmentations are included in the  calculations.
Results for the
calculations with only the binary collisions are also given (diamonds).
Lower frame : The multiplicity density of photons having $p_T >$ 2 GeV.}
  \label{fig2}
  \end{center}
\end{figure}

We next allow for fragmentation processes after binary interactions as 
discussed above. The lower set of points in the upper frame of 
Fig.~\ref{fig2} (squares) includes only QCD induced fragmentations 
of partons. A comparison with Fig.~\ref{fig1} (diamonds) shows that 
this already leads to an increase in the photon emission by about 
a factor 4. The reason for this surprisingly dramatic
effect is that the fragmentation leads to a large multiplication of 
partons and thus to an enhanced photon yield to the increase in the
number of multiple scatterings. In comparison, the increase in the
photon yield due to the fragmentation of photons off the scattered 
quarks (triangles) is small, except below 1 GeV, filling up the 
artificial gap caused by the cut-off $p_T^{\text{min}}$ in the binary
collisions.  The strong enhancement in the low $p_T$ region is not 
surprising, since radiative processes generally favor low-energy final 
states.

The lower frame of Fig.~\ref{fig2} shows the rapidity distribution 
of photons with $p_T > 2$ GeV for the same two cases as are discussed
in the upper frame, but includes also the distribution obtained when
only binary interactions are allowed (diamonds, corresponding to the
uppermost curve in Fig.~\ref{fig1}). The enhancement of photon 
production due to the inclusion of fragmentation processes occurs 
mainly around mid-rapidity, where a dense partonic system is created
and most of the fragmentation processes contribute.  We chose the
cut-off $p_T > 2 p_T^0$, because the photon yield below this value 
is expected to receive a large contribution from hadronic reactions,
which do not interest us here.

The strength of the source of single photons is often discussed in terms
of the ratio $(\gamma/\pi^0)$, as most of the back ground to 
them comes from the decay of $\pi^0$.  We 
can not get a precise value for this unless we supplement our description
with a model for the hadronization. A reasonable alternative is provided
by the ratio of ($\gamma/$partons) as the multiplicity of pions will be 
proportional to the  multiplicity of partons. We find that this quantity is
of the order 0.04\% when only multiple scatterings among partons
is considered and rises to 0.15\% when fragmentations are allowed and
which considerably enhance the multiple scatterings.

The results shown so far do not account for the possible LPM 
suppression of radiation from the scattered partons. An accurate 
implementation of this effect in a semi-classic treatment is difficult. 
To explore the possible magnitude of this effect, we consider here
a treatment in which the scattered parton, having energy $E_a$ and 
time-like virtuality $m_a^2$, is forbidden to fragment if it re-scatters 
before the formation time of the radiated parton- $E_a/m_a^2$ has 
expired. This ansatz should provide us with an upper limit of the 
suppression as it does not account for the relative phases of the 
two scatterings. We find that the emission of photons at about 
2 GeV is reduced by a factor of 3.5 if the LPM suppression is treated 
in this manner. Most of the enhancement due to fragmentation 
processes visible in Fig.~\ref{fig2} disappears in this limit.
Also, increasing the pQCD cut-off $(p_T^{\text{min}})^2$ to 1 GeV$^2$ 
reduces the number of binary collisions and the build-up of cascading 
activity. The net result is a decrease in the photon yield by a factor 
of about 1.5 beyond $p_T = 1$ GeV near central rapidities.

\begin{figure}[tb]
  \begin{center}  
  \epsfig{file=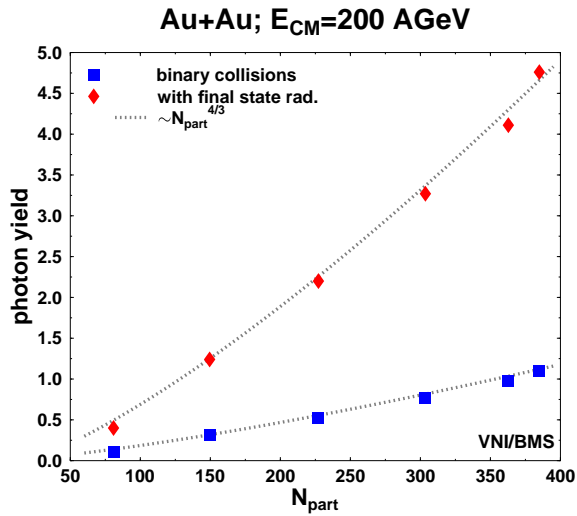,width=8.2cm}
  \caption{Integrated photon multiplicity as a function
of $N_{\rm part}$. Squares denote a calculation with only binary 
parton-parton collisions whereas diamonds show a calculation including
fragmentation processes. The scaling with $N_{part}^{4/3}$ is indicative
of the photon yield being directly proportional to the number of hard 
collisions.}
  \label{fig3}
  \end{center}
\end{figure}

How does the photon production scale with the number of participant 
nucleons $N_{\rm part}$? Figure~\ref{fig3} shows the total integrated 
photon yield (neglecting hadronic and thermal contributions)
as a function of $N_{\rm part}$. The squares denote a calculation 
without fragmentation processes, whereas the diamonds show the 
full calculation, including fragmentation. For both calculations we 
observe a scaling of the photon yield with $N_{\rm part}^{4/3}$ 
(dotted curves), suggestive of a scaling with the number of binary 
parton-parton collisions~\cite{Fei:76,Ha:82} rather than participant
number. This is a valuable result, since it provides a direct connection
between a measurable quantity (the photon yield) and the unobservable 
number of hard collisions during the time-evolution of the reaction. 
It will be interesting to see in future studies whether the production 
of heavy quarks and the suppression of high-$p_T$ hadrons reflect
similar scaling properties.

In our calculation we have used lowest order pQCD matrix elements to 
study the collisions. One could in principle use a $K$-factor to 
account for higher order effects. We have not done so here, as the 
parton shower mechanism for the final state radiations amounts to 
the inclusion of collinear emissions to all orders.  At least for 
photon production in a QGP, rates are now available up to two 
loops~\cite{au:98}, and to complete leading order~\cite{ar:01}.
It is not clear how these calculations can be incorporated in the 
PCM, unless a (local) temperature is associated with the system.  
Our current calculation neglects medium modifications of the matrix 
elements and fragmentation functions.  We intend to study these 
effects in a forthcoming publication.

In summary, we have calculated the production of high energy photons 
in collision of gold nuclei at 200 A GeV in the Parton Cascade Model.
Multiple scattering of partons is seen to lead to a substantial 
production of high energy photons, which rises further when parton 
multiplication due to final state radiation is included.
The photon yield is found to scale as $N_{\text{part}}^{4/3}$ and
is directly proportional to the number of hard parton-parton collisions
in the system, providing valuable information on the pre-equilibrium 
reaction dynamics of the system.
Our calculations suggest that that at the much higher energies to be 
reached at the Large Hadron Collider, the parton cascades developing
in the nuclear collision will produce an even larger ``firework''
of photons.

\begin{acknowledgments}  
This work was supported in part by DOE grants DE-FG02-96ER40945 and
DE-AC02-98CH10886 and the Natural Sciences and Engineering Research
Council of Canada. 
\end{acknowledgments}

\end{document}